\documentclass[preprint2,longabstract]{aastex}
\usepackage[]{natbib}
\newcommand{\lya}{Ly$\alpha$}
\newcommand{\flya}{$f_{\mathrm{Ly}\alpha}$}
\newcommand{\flyapos}{$f_{\mathrm{Ly}\alpha}^+$}
\newcommand{\lyam}{{\rm Ly}\alpha}
\newcommand{\ha}{H$\alpha$}
\newcommand{\ham}{{\rm H}\alpha}
\newcommand{\hb}{H$\beta$}
\newcommand{\hi}{H{\sc i}}
\newcommand{\hii}{H{\sc ii}}

\newcommand{\ebv}{$E(B-V)$}
\newcommand{\msun}{{\cal M}$_\odot$}

\newcommand{\sbs}{SBS\,0335--052}
\newcommand{\eso}{ESO\,338--04}
\newcommand{\ngc}{NGC\,6090}
\newcommand{\tol}{Tol\,65}
\newcommand{\haro}{Haro\,11}
\newcommand{\iras}{IRAS\,08339+6517}
\def\mathbi#1{\textbf{\em #1}}

\slugcomment{Accepted by the Astronomical Journal}

\shorttitle{Lyman Alpha Atlas of Local Starbursts}
\shortauthors{\"Ostlin, Hayes, Kunth et al.}

\begin{document}

\title{The Lyman alpha morphology of local starburst galaxies: 
		release of calibrated images }

\author{G\"oran \"Ostlin\altaffilmark{1}}
\affil{Department of Astronomy, Stockholm University, AlbaNova University Center, 10691 Stockholm, Sweden}
\email{ostlin@astro.su.se}
\author{Matthew Hayes}
\affil{Geneva Observatory, University of Geneva, 51 chemin des Maillettes, 1290 Sauverny, Switzerland}
\email{matthew.hayes@unige.ch}
\author{Daniel Kunth}
\affil{Institut d'Astrophysique de Paris, Paris (IAP), 98 bis boulevard Arago, 75014 Paris, France}
\author{J.~Miguel Mas-Hesse\altaffilmark{2}}
\affil{Centro de Astrobiolog{\'a} (CSIC-INTA), E28850 Torrejon de Ardoz, Madrid, Spain}
\author{Claus Leitherer}
\affil{Space Telescope Science Institute, 3700 San Martin Drive, Baltimore, MD 21218, US}
\author{Artashes Petrosian}
\affil{Byurakan Astrophysical Observatory and Isaac Newton Institute of Chile,
	Armenian Branch, Byurakan 378433, Armenia}
\and
\author{Hakim Atek}
\affil{Institut d'Astrophysique de Paris, Paris (IAP), 98 bis boulevard Arago, 75014 Paris, France}

\altaffiltext{1}{Oscar Klein Centre for Cosmoparticle physics, Department of Astronomy, Stockholm  University, Sweden}
\altaffiltext{2}{Laboratorio de Astrof{\'i}sica Espacial y F{\'i}sica Fundamental
(LAEFF-INTA), POB 78, Vva. Canada, Spain}

\begin{abstract}

We present reduced and calibrated high resolution Lyman-alpha (\lya) images for a 
sample of six local star forming galaxies.
Targets were selected to represent a range in luminosity and metallicity and to 
include both known \lya\ emitters and non-emitters.
Far ultraviolet imaging was carried out with the Solar Blind Channel of the 
Advanced Camera for Surveys on Hubble Space Telescope in the F122M 
(\lya\ on-line) and F140LP (continuum) filters.  
The resulting \lya\ images are the product of careful modeling of both 
the stellar and nebular continua, facilitated by supporting HST
imaging at $\lambda \approx 2200$, 3300, 4400, 5500, \ha\ and 8000\AA, 
combined with {\em Starburst\,99} evolutionary synthesis models, and
prescriptions for dust extinction on the continuum.  
In all, the resulting morphologies in \lya, \ha, and UV-continuum are
qualitatively very different and we show that the bulk of \lya\ emerges in 
a diffuse component resulting from resonant scattering events.
\lya\ escape fractions, computed from integrated \ha\ luminosities and 
recombination theory, are found never to exceed 14\%. 
Internal dust extinction is estimated in each pixel and used to correct 
\lya\ fluxes. 
However, the extinction corrections are far too small (by factors from 2.6 to infinity)
 to reconcile the emerging 
global \lya\ luminosities with standard recombination predictions.   
Surprisingly, when comparing the global equivalent widths of \lya\ and \ha, 
the two quantities appear to be {\em anti-correlated}, which may be due to the
evolution of mechanical feedback from the starburst. 
This calls for caution in the interpretation of  \lya\ observations 
in terms star formation rates.
The images presented have a physical resolution 3 orders of magnitude better 
than attainable at high redshifts from the ground with current instrumentation and 
our images may therefore serve as useful templates for comparing with observations
and modeling of primeval galaxy formation. 
We therefore provide the reduced \lya, \ha, and continuum images to the community.

\end{abstract}


\section{Introduction}

The search for galaxies experiencing their first phase of violent star
formation is one of the great challenges for observational cosmology.
In this context, the importance of the Lyman-alpha emission line (\lya) can 
hardly be overstated. 
Young galaxies with low dust content should produce strong \lya \
emission with equivalent widths, $W_{\lyam}$, larger than  100\AA\
\citep{Charlot_Fall93,Schaerer03}.
With a rest wavelength of 1216~\AA\ the \lya\ line is observable
from the ground in the
optical window at redshifts $z\sim 2$ to 6, and in principle in the near
IR as far as $z\sim 20$.

Significant samples of galaxies can currently be established up to 
$z > 5$
using continuum (Lyman break, e.g. 
\citealt{Iwata03,Stanway03,Bremer_Lehnert04,Shimasaku05})  
or emission line (\lya, e.g.
\citealt{Rhoads01,Hu04,Venemans04,Ouchi05})
techniques. 
However, while there is now much information regarding the continuum
morphologies of galaxies and their evolution with redshift 
\citep[e.g.][]{Windhorst02,Lotz06,Ravindranath06,Law07},
the situation is very different for \lya\ where morphologies are all but
unknown. 
At high-$z$ ($\gtrsim 2$) when \lya\ is observable from the ground, even the best 
seeing conditions result in observations with a spatial sampling of several kpc. 
At $z<2$, space-based observation becomes a requirement, although still very few 
instruments have been capable of emission line imaging in the UV. 
In the current study we take a first step towards remedying the almost
complete lack of information 
regarding \lya\ morphologies of local star forming galaxies.

First theoretical models proposed that primeval galaxies should be
strong UV emitters both in continuum and line emission
\citep{Partridge_Peebles67,Meier76,Sunyaev78}, 
although early observations proved disappointing.
However, at these times no high-$z$ normal galaxies were known and the UV domain 
was terra incognita.
An early attempt to study \lya\ emission from three local star forming galaxies 
using the International Ultraviolet Explorer (IUE) found  that \lya\ emission was
only present in the most metal-poor galaxy studied \citep{Meier_Terlevich81}. 
Subsequent studies of larger samples 
\citep{Charlot_Fall93}
confirmed a weak correlation of \lya\ and the metallicity of the gas; not an 
unpredicted result given that the dust content generally correlates with metallicity. 
However, corrections for dust were found {\em not} to reconcile the \lya/\hb\ line
ratios with those expected from recombination theory 
\citep{Giavalisco96}
and it became clear from these early studies that some other processes must be at
work. 
Indeed, resonant scattering of \lya\ in \hi\ had long been studied theoretically
\citep{Osterbrock62,Adams72,Harrington74}.
The sub-recombination line ratios and low $W_{\lyam}$ can be explained by
multiple resonant scatterings locally trapping \lya\ photons, redirecting them
and causing them to traverse much greater integrated path lengths than
non-resonant continuum or emission lines. 
Thus, even minute quantities of dust may result in significant or complete
attenuation of \lya\
\citep{Neufeld90}.

A major breakthrough in the detection of significant numbers of LAEs at high-$z$ 
\lya\ came with 
\cite{Cowie_Hu98}, 
and today \lya\ emission is arguably the most successful method for examining the
galaxy population at the highest redshifts. 
Probing intrinsically fainter objects than the Lyman break technique, \lya\ 
emission serves either as the spectral signature through which
to identify high-$z$ galaxy candidates or the definitive redshift
identification of candidate objects. 
The impact on astrophysical cosmology has been enormous with \lya\ surveys
being used to study star-formation rates
\citep{Kudritzki00,Fujita03,Gronwall07}, 
large-scale structure
\citep{Venemans04,Ouchi05,Nilsson07}, 
test the ionization fraction of the intergalactic medium
\citep{Kashikawa06,Dijkstra07}, 
and identify potential sites of population {\sc iii} star formation
\citep{Malhotra_Rhoads02,Shimasaku06}. 

While it's difficult to argue against the impact of the high-$z$ \lya\ studies,
some inconsistencies call for caution. 
Most notably that star-formation rates (SFR) in individual galaxies are often found 
to be lower when estimated by \lya\ when compared with UV continuum 
\citep{Ajiki03,Taniguchi05}. 
Underestimates of the cosmic SFR density also tend to be underestimated by
larger values when measured from LAEs
\citep{Rhoads03,Kodaira03}. Moreover, LAEs have an inherent selection bias since
only sources which show emission are included. 
In the sample of $\sim 800$ UV selected galaxies at $z\sim 3$,
(also known as Lyman Break Galaxies (LBG) 
\citealt{Shapley03})
found a median $W_{\lyam}$ of $\sim 0$\AA,  with only $\sim 25$\%
showing $W_{\lyam} \ge 20$\AA.
How these values evolve towards fainter continuum levels is unknown.
Furthermore
\citep{Ouchi03} 
noticed that bright LAEs appear more strongly biased against the underlying dark
matter halo distribution than continuum-bright sources. 
While the semi-analytical models for the space density of LAEs seem to be
converging with the observed values
\citep{LeDelliou05}, 
uncertainties are still present and
therefore open to discussion. 
In order to derive robust quantities from the LAE population, it is vital to
understand the origin of these biases. 

Again, results from low-$z$ imply that the interpretation of \lya\ observation 
at high-$z$ may not be so clear-cut. 
High spectral resolution studies with HST showed the \lya\ properties of 
local starbursts to be complex, in realizing that a low dust content is by no 
means a guarantee for observing \lya\ in emission. 
Again, little correlation is found between \lya\ emission or absorption and dust or 
metallicity
\citep{Kunth94,Lequeux95,Thuan_Izotov97}.
\cite{Kunth98} 
clearly demonstrated that when \lya\ is seen in emission it is systematically
P\,Cygni in form and associated with signatures of an outflowing neutral ISM, 
while starbursts showing \lya\ absorption features show no such outflow signatures. 
These results, also noticed at high-$z$ in LBG spectra
\citep[e.g.][]{Shapley03}, 
have profound implications for the escape of \lya\ from starbursts: they suggest
an outflowing medium to be a requirement in order for photons to avoid resonant trapping in
the neutral medium. 
Hydrodynamic models 
\citep{Tenorio-Tagle99}
predict an evolutionary sequence for the \lya\ profile as a function of the mechanical
energy return from the starburst, demonstrating how absorption, emission and 
P\,Cygni phases may all arise during different phases. 
\cite{Mas-Hesse03} 
used this physical evolutionary scenario to interpret the range of \lya\ profiles 
observed at low-$z$ as an evolutionary sequence in the starburst population. 
To get
a feeling for the astrophysics involved, we 
encourage the reader to look at Fig. 18 in \cite{Mas-Hesse03}.

Much theoretical effort has gone into the study of \lya\ line profiles. 
\cite{Ahn04} 
and 
\cite{Verhamme06} 
have examined the formation of P\,Cygni profiles from Lyman break-type galaxies,
finding secondary emission peaks redwards of the P\,Cygni emission feature resulting 
from one or more backscatterings in the expanding neutral medium.
\cite{Neufeld91}
and 
\cite{Hansen_Oh06}
have examined the transfer of \lya\ in a multiphase ISM. 
For models in which dust is embedded in dense \hi\ clouds, they find that \lya\
photons can in fact preferentially avoid dusty regions and escape having
experienced lower total dust optical depths than non-resonant photons.
Thus equivalent widths may be artificially boosted, potentially causing ordinary 
star-forming objects to appear as more exotic systems. 
	 
What the low-$z$ spectroscopic studies have not been able to accomplish is 
examination of global values relating to \lya: spectroscopic observations miss the
fraction (often the majority) of light which comes from outside the aperture or slit. 
This is highly likely to be the case for \lya\ where resonant scattering may
cause decoupling of \lya\ from other wavelengths and \lya\ photons
may scatter in the galaxy and emerge far from their sites of production. 
Ionized holes or porosity in the ISM may also increase the chance of \lya\ escape along
certain paths. 
Ultimately, there is little evidence for the spatial correlation between \lya\
emission and UV continuum at all. 
These considerations were the motivation for our pilot low-$z$ imaging study of
six local star-forming galaxies using the Advanced Camera for Surveys (ACS) 
onboard HST. 
The intent was to map \lya\ morphology and study the effects of dust,
kinematics, and luminosity on the escape of \lya\ and the sample was hand-picked
to reflect a range of these parameters.
First results were presented in 
\cite{Kunth03}.

Besides \lya, images have been acquired in \ha\ and six continuum bandpasses between
1500\AA\ and the I-band, enabling not only a comparison between \lya\ and
\ha, but also detailed modeling of the stellar population and dust. 
Detailed studies have been presented for two galaxies in the sample, both
low-metallicity UV-bright starbursts, \eso\ and \haro\
\citep{Hayes05,Hayes07}.
These studies found that emission and absorption varied on
small scales in the central starburst regions, exhibiting very little
correlation with the morphology probed by stellar continuum observations. 
Damped \lya\ absorption was found in front of many UV-bright, young star
clusters that must be a significant source of ionizing photons.
Using \ha, it was found that when \lya\ does emerge from central regions, it
is typically at surface brightness well below the value predicted by recombination 
theory. 
In contrast, large low surface brightness \lya\ halos were found surrounding the 
central regions where typically \lya/\ha\ ratios and equivalent widths exceed
those predicted for recombination.
Furthermore, modeling of the stellar population revealed that \lya\ appears to emerge 
from regions either too old to produce ionizing photons and/or too dusty to permit the
transfer of \lya.
Synthesis of these results points to the resonant scattering process being
highly significant in the morphological structure. 

It should be noted that these HST imaging observations are the only
ones in existence that allow the comparison of \lya\ with \ha\ 
and stellar continuum and that they have a physical resolution, in the most
distant case, 2 orders of magnitude better than that attainable at high-$z$ under
the best observing conditions. 
These data have now been fully processed, and continuum subtracted. 
Details of the continuum subtraction methodology can be found in \cite{Hayes09}, 
hereafter referred to as Paper I. 
In this article we discuss the \lya\ line morphology, comparing it against \ha\
and stellar continuum, and compute integrated fractions of escaping \lya\
photons. 
The resolution of the data is so high that we believe them to be of interest for 
comparison with high-$z$ samples of \lya-emitting galaxies and also for the computational 
modeling of \lya\ radiative transfer. 
As a service to the community we release the processed data on the 
{\em Centre de Donn{\'e}es astronomiques de Strasbourg (CDS)}.

The article proceeds as follows: 
in Sections~\ref{sect:observations} and ~\ref{sect:processing} we present the observations and data
processing methods; 
in Section~\ref{sect:results} we present the results, which we discuss in 
Section~\ref{sect:discussion}; 
Section~\ref{sect:perspectives} is dedicated to future perspective and
possibilities;  
in Section~\ref{sect:conclusions} we present our concluding remarks and  
describe how to retrieve the released data. 

Throughout this paper we will adopt $H_0=72$~km~s$^{-1}$~Mpc$^{-1}$ for the
Hubble constant.

\section{Observations} \label{sect:observations}

Six galaxies were selected for observations with HST. 
Since this is a pilot study, our sample is hand-picked.
We aimed to provide a reasonable coverage of the 
starburst luminosity, metallicity and dust content. 
Four known \lya\ emitters were included from the samples of 
\cite{Calzetti_Kinney92}
and 
\cite{Kunth98}: \haro, \iras, \ngc, and  \eso; 
and two non-emitters from 
\cite{Thuan_Izotov97}: \sbs\ and \tol.
Since \lya\ regulation is  affected by ISM kinematics, selection
of emitting targets was also based on the kinematic information imprinted on the
observed spectroscopic line profile 
\citep{Mas-Hesse03}. 
Our sample is summarized in Table \ref{targets}.

\begin{table*}[th]
\scriptsize
\begin{center}
\caption{Target summary\label{targets}}
\begin{tabular}{llrrccccccc}
\tableline\tableline  \\
Target & Alternative & RA(2000) & Dec(2000) &$E(B-V)_{\rm MW}$ & log($n_{\rm H I}$)$_{\rm MW}$ & $v_r$ & $12+$ & $M_B$ & Em/Abs$^{a}$ & Ref \\
name &  name &  & & &  & (km/s) & $\log({\rm O/H})$ &  \\  \\
\tableline  \\
\haro\ & ESO\,350--38   & 00:36:52.5  &  --33:33:19  & 0.049 & 20.4 & 6175 & 7.9 & --20 & Em  & 1 \\
\sbs\  & SBS\,0335-052E & 03:37:44.0  &  --05:02:40 & 0.047 & 20.6 & 4043 & 7.3 & --17 & Abs & 2 \\
\iras\ & PGC\,024283    & 08:38:23.2  &  +65:07:15  & 0.092 & 20.6 & 5730 & 8.7 & --21 & Em  & 3 \\
\tol\  & ESO\,380--27   & 12:25:46.9  &  --36:14:01 & 0.074 & 20.7 & 2698 & 7.6 & --15 & Abs & 4 \\
\ngc\  & Mrk\,496       & 16:11:40.7  &  +52:27:24  & 0.020 & 20.2 & 8785 & 8.8 & --21 & Em  & 3 \\
\eso\  & Tol\,1924-416  & 19:27:58.2  &  --41:34:32 & 0.087 & 20.7 & 2832 & 7.9 & --19 & Em  & 1 \\  \\
\tableline
\end{tabular}
\tablenotetext{a}{Em: known \lya-emitter, Abs: known \lya -absorber }
\tablecomments{$E(B-V)_{\rm MW}$ is the Galactic extinction according to
\cite{Schlegel98} and log($n_{\rm H I}$)$_{\rm MW}$ is the Galactic \hi\ 
column density along the line of sight.
Reference refers to nebular metallicity measurements:
1: \cite{Bergvall_Ostlin02},
2: \cite{Papaderos06},
3: \cite{Gonzalez-Delgado98},
4: \cite{Izotov01} }
\end{center}
\end{table*}

The targets were observed with the Advanced Camera for Surveys (ACS)
onboard the Hubble Space Telescope under two general observer programs: 
GO\,9470 for acquisition of far UV data in the F122M and F140LP 
filters using the solar blind channel (SBC); and GO\,10575 
for obtaining \ha, near-UV, and optical broad-band observations with 
the High Resolution (HRC) and Wide Field (WFC) cameras. 
The F122M observations were obtained during the `SHADOW' part of the orbit in order 
to reduce the geocoronal \lya\ background.
In addition some
complementary archival WFPC2 data was used. 
The observations are presented in Table~\ref{obs}.

\begin{table*}[th]
\scriptsize
\begin{center}
\caption{Observations: HST filter names, central wavelengths, and exposure times in seconds.}\label{obs}
\begin{tabular}{lcccccccc}
\tableline\tableline  \\
Target / Filter &  F122M & F140LP & F220W & F330W & F435W & F550M & FR656N & F814W \\ 
$~~~~\lambda_{\rm central}$~(\AA )       & 1274 & 1527 & 2257&  3362&  4312&  5582&  variable& 8121\\ 
\tableline  \\
\haro\ & 9095 & 2700 & 1513~ & 800~ & 680~ & 471~ & 680~ & 100 \\
\sbs\  & 9000 & 2700 & 1660~ & 800~ & 680~ & 430~ & 680~ & 4400$^{a}$ \\
\iras\  & 9000 & 3000 & 1590~ & 730~ & 360$^{c}$ & 560$^{c}$ & 1024$^{c}$ & -- \\
\tol\ & 9095 & 2700 & 1728~ & 800~ & 680~ & 498~ & 680$^{d}$ & 3400$^{e}$ \\
\ngc\  & 9095 & 3000 & 1683~ & 800~ & 680~ & 645~ & 680~ & 100 \\
\eso\  & 9000 & 3000 & 1800$^{b}$ & 900$^{b}$ & 800$^{b}$ & 830~ & 1214~~ & ~520$^{b}$ \\ \\
\tableline
\end{tabular}
\tablenotetext{a}{Archival WFPC2/WF3/F791W image originally obtained in program 5408 (PI: Thuan)}
\tablenotetext{b}{Archival WFPC2/PC data (F218W, F336W, F439W \& F814W) from program 6708 (PI: \"Ostlin)}
\tablenotetext{c}{Data obtained with ACS/HRC}
\tablenotetext{d}{Data not useful}
\tablenotetext{e}{Archival WFPC2/WF3/F814W image originally obtained in program 6678 (PI: Thuan)}
\tablecomments{ 
Unless  noted otherwise the F220W, F330W, F814W were obtained with ACS/HRC; and F435W, F550M, FR656N with ACS/WFC}
\end{center}
\end{table*}

\section{Data reduction and processing} \label{sect:processing} 

All\footnote{With the exception of \tol\ due to guide star problems, see Sect. 4.5} 
pipeline calibrated data from the different cameras were drizzled 
to the same scale 
(0.025\arcsec/pixel) and orientation using the task {\sc multidrizzle}  
in {\sc pyraf}. Remaining pixel shifts were fixed using the  {\sc geomap} 
and {\sc geotran}, and any remaining cosmic rays were removed using 
{\sc credit}.
Next, all the images for a particular target were convolved to a common point
spread function (PSF) using the image with the broadest PSF as a reference. 
This in general was F550M, except for 
\sbs\ and \tol\ where the archival I-band data had been obtained with the WF3 chip 
of WFPC2. 
Each image was then sky subtracted using {\sc fit/flat\_sky} in
{\sc midas} using a low order (0 or 1) polynomial and a common sky region 
for the different passbands. 

Obtaining \lya\ line-only images has been shown to be a rather involved
process, requiring the use of spectral model fitting, and is presented in
Section~\ref{sect:contsub}. 
Model parameters selected for the fitting are discussed in
Section~\ref{sect:sb99pars}.
Fitting in turn requires a threshold signal--to--noise ratio to yield tolerable
results, requiring binning of the data, discussed in 
Section~\ref{sect:binning}.

\subsection{Continuum subtraction}\label{sect:contsub}

Going from F122M and F140LP to continuum subtracted \lya \ images is 
non-trivial for a number of reasons: the on-line filter is quite
broad with a red wing, the effective wavelength of the off-line filter 
is rather far to the red side of the \lya \ line, neutral hydrogen in
the Milky Way absorbs part of the flux in the F122M filter, and the UV
continuum is less well described by a power law near \lya \ than at
slightly longer wavelengths. In effect, adopting a simple assumption 
of a UV power-law ($f_\lambda \propto \lambda^\beta$) and neglecting 
the other effects mentioned can  produce results ranging 
from strong \lya \ emission to damped absorption simply by varying 
$\beta$. 
Indeed, \cite{Hayes_Ostlin06}
have shown that also for high-$z$ imaging observations, estimation of the
continuum at \lya \ may be non trivial.

To overcome these limitations we have developed 
a method for continuum subtraction of ACS/SBC \lya\ images. It is described in 
detail in \cite[][hereafter Paper {\sc i}]{Hayes09}  where we also perform thorough tests of the 
methodology and parameter dependencies. 
In short, the method  utilizes {\em Starburst\,99} \citep[][hereafter SB99]{Leitherer99} spectral
evolutionary synthesis models and a set of UV and optical continuum band passes plus \ha\ to 
perform a spectral energy  distribution (SED) fit of up to two stellar and one nebular 
components, with reddening treated as a free parameter. 
The output is the {\em Continuum Throughput Normalization (CTN)} factor which
determines how to scale the F140LP flux in order to accurately continuum subtract
the F122M images. 
 
The CTN is primarily dependent on the age and reddening of the stellar population
dominating the UV flux. By sampling the UV spectral slope longwards of 1500\AA\
and the 4000-break a non degenerate solution for CTN is obtained. Additional 
optical bandpasses and \ha\ are included to constrain the  contribution from 
an underlying older stellar population and  ionized gas:
Determining the continuum level at \lya\  depends on fitting the stellar 
population age, which  is sensitive to the strength of the 4000\AA\ break. 
Contamination  of this feature by an aged population that does not contribute 
to the continuum at 1216 \AA\ could therefore result in erroneous age fits and 
estimates of CTN. Hence the need for a 
two population SED fit. The philosophy is similar regarding the Balmer jump 
which may be significant where \hii\ shells or filaments are well resolved, 
and is accounted for by treating nebular continuum independently.

The \ha\ images have been continuum subtracted within the SED fitting
software and have been corrected for the contribution from [N{\sc ii}]
using literature data.

For all targets but two we have used two stellar plus one nebular component
in the continuum subtraction (this is referred to as method\,{\sc v} in 
Paper\,I). For \iras\ and \tol\ we did not have sufficient data for this 
purpose and used an SED fit with fewer components (see Sect. \ref{individuals}).

\subsection{{\em Starburst\,99} and continuum modeling parameters}\label{sect:sb99pars}

For this paper we use a standard setup of model parameters:
A 
\cite{Salpeter55}
IMF ($\alpha = 2.35, dN \propto M^{-\alpha} dM$) with
lower and upper mass limits of 0.1 and 120\msun, respectively. 
To model the starburst and underlying stellar populations we use instantaneous
burst (single stellar populations); Geneva tracks with enhanced mass loss, and 
Pauldrach/Hillier atmospheres for the UV were used throughout.  
While derived ages and reddenings are more sensitive to assumed model 
parameters, the derived CTN and hence \lya\ flux is indeed much less sensitive
\citep[see][]{Hayes05,Hayes09}.  
In general, the most important concerns are the use of correct metallicity and 
extinction law. 
For all galaxies we have adopted the SMC reddening law 
\citep{Prevot84} 
since the intense UV radiation field of these galaxies tends to destroy the 
graphite component of the dust, leading to smooth extinction laws without the 
2175~\AA\ feature, and thus providing consistently improved fits over other laws 
\citep[e.g.][]{Mas-Hesse_Kunth99}.
The effect of changing this is discussed in detail in Paper {\sc i}. 

For metallicity, we opt for the metallicity in {\em Starburst\,99} closest to 
the observed nebular oxygen abundance (see Table 
\ref{targets}).
Hence for \iras\ and \ngc\ we used $Z=0.02$, for \haro\ and \eso\ we used 
$Z=0.004$, and for \sbs\ and \tol\ we used $Z=0.001$. 
Using Monte Carlo simulations (Paper {\sc i}) we show that the accuracy of the 
continuum subtraction is not greatly affected, provided the assumed 
metallicity is within 50\% of the true value. 
For all galaxies and metal abundances presented here, that is indeed the case.

\subsection{Binning}\label{sect:binning}

For the continuum subtraction to yield photometrically reliable results a
signal--to--noise (S/N) in each passband of between 5 and 10 is required
\citep{Hayes09}.
This requires some smoothing or binning of the data in the low surface 
brightness regions. In order to simultaneously preserve the spatial 
information in the high S/N central regions, adaptive binning 
was used. 
We here make use of the Voronoi binning code of 
\cite{Diehl_Statler06}
which is a generalization of the binning algorithm of 
\cite{Cappellari_Copin03}.
We use the F140LP image and its variance map to bin together pixels to
conglomerate groups (known as `spaxels') to improve S/N locally, while
conserving surface brightness. 
Each spaxel accretes pixels until a S/N of 10 has been attained or until
the maximum spaxel size of 1600 pixels (1 $\square\arcsec$) was reached. 
The same binning pattern was then applied to the other passbands.

\subsection{Uncertainties}\label{sect:errors}

The accuracy of the continuum subtraction as a function of the input
parameters is addressed in 
\cite{Hayes09}, 
where noise is added to
idealized observations using Monte-Carlo simulations. 
In that article we consider only idealized, model data.
These Monte Carlo simulations found the uncertainty (including the
{\em statistical} uncertainty arising from noise in our fitting procedure, and 
{\em systematic} uncertainty accounting for our own estimation of model
parameters)
on the \lya\ flux in a given spaxel to be a function of equivalent width. 
In general this has been shown to be below 30\%  for EW(\lya)=10~\AA\ and
S/N=10 (the fiducial minimum value set by the binning, 
Section~\ref{sect:binning}), with the error decreasing with increasing 
EW(\lya) and S/N.

In this paper we have in addition performed Monte-Carlo simulations to
assess the statistical uncertainties on the real data: 
to each spaxel, random errors have been added in accordance with the 
photometric uncertainties for each passband. 
For each galaxy 100 Monte Carlo realizations were performed.
Thus, the 8 individual science images for each object were regenerated 100 times 
with each spaxel randomly replaced by a Gaussian variate of the original, in
accordance with its error.
From these we created 100 new continuum subtracted images for each target,
retaining all the important statistics of the spaxels in the final frame. 
We conclude that the statistical errors on the {\em integrated} \lya\ 
properties in Table~\ref{tab:results2} are very small, ranging from 4\% in the 
case of \tol\ to 0.5\% for \eso. 
The statistical S/N for each individual spaxel is of course smaller in 
general: the luminosity weighted averages range from 
$\langle S/N\rangle _L=0.26$ for \tol\ to 4.9 for \haro. 
Statistical uncertainties on \ha\ properties are in general orders of magnitude
smaller.

A larger source of uncertainty is likely to be errors that result from sky 
subtraction and flat fielding. 
The uncertainty of the sky level were, for each passband and galaxy, again fed 
into a Monte-Carlo simulation in a similar manner to that described above. 
A very conservative result is that flatfield and sky subtraction residuals can 
at most cause global errors on the 20\% level. Since these kind of errors, 
when present, tend to have low spatial frequency they are likely to affect the 
global levels rather than causing pixel--to--pixel variations.
 
In addition we have estimated the uncertainties by calculating the standard 
deviation of the background of the resulting \lya\ level outside the 
galaxies, at a scale of 1 $\square\arcsec$, i.e. the maximum spaxel 
size (see Table~\ref{tab:results2}).

\section{Results} \label{sect:results}

Various analyses are performed on the processed images: direct photometric
analysis and comparison with model-determined properties.
We examine both small-scale, local relationships and global quantities, and  
our results are presented in a number of Figures and Tables. 
More detailed analysis for individual targets have either been presented
previously \citep[][]{Hayes05,Hayes07} or will be the scope of subsequent papers.

\subsection{Morphologies}

\begin{figure*}
\begin{center}
\includegraphics[width=3.2cm]{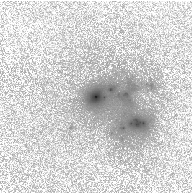}
\includegraphics[width=3.2cm]{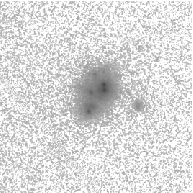}
\includegraphics[width=3.2cm]{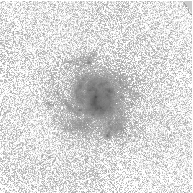}\\
\includegraphics[width=3.2cm]{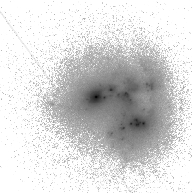}
\includegraphics[width=3.2cm]{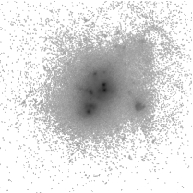}
\includegraphics[width=3.2cm]{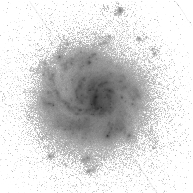}\\
\includegraphics[width=3.2cm]{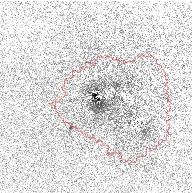}
\includegraphics[width=3.2cm]{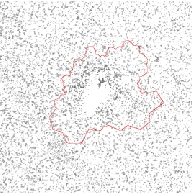}
\includegraphics[width=3.2cm]{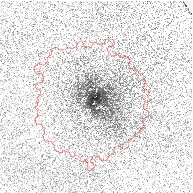}\\
\includegraphics[width=3.2cm]{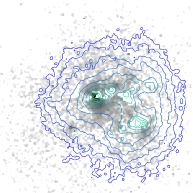}
\includegraphics[width=3.2cm]{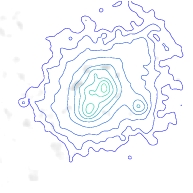}
\includegraphics[width=3.2cm]{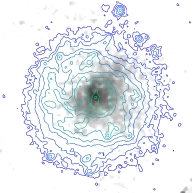}\\
\includegraphics[width=3.2cm]{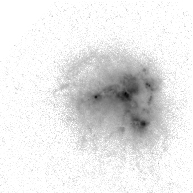}
\includegraphics[width=3.2cm]{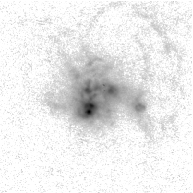}
\includegraphics[width=3.2cm]{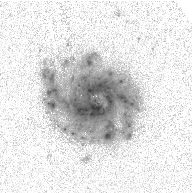}
\end{center}
\caption{Greyscale images.
From {\em top} to {\em bottom}: Reduced F122M and F140LP;
continuum subtracted \lya\ resulting from unbinned images 
but with the $\mu_{AB,1500}=23$ mag/$\square\arcsec$ 
isophotal mask generated from binned data overlaid; 
adaptively filtered \lya\ with F140LP isophotes overlaid; 
and continuum subtracted unbinned \ha. 
The isophotal mask from the binned data is used later for quantitative 
photometry. 
The FUV contours show 25~mag/$\square\arcsec$ as the faintest isophote, with
each subsequent isophote representing an increase by 1 mag/$\square\arcsec$.
All intensity scaling is logarithmic. North is up, East to the left.
Images are square cutouts with the length of one side in arcsec given in 
parentheses.
From {\em left} to {\em right}: 
\haro\ (20\arcsec) , \sbs\ (10\arcsec) and \iras\ (20\arcsec). 
In some images residual features from the ACS/SBC detector gap are visible.
These areas have not been included in the photometry.}
\label{grey1}
\end{figure*}

\begin{figure*}
\begin{center}
\includegraphics[width=3.2cm]{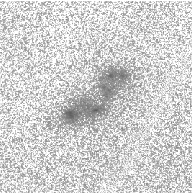}
\includegraphics[width=3.2cm]{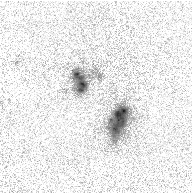}
\includegraphics[width=3.2cm]{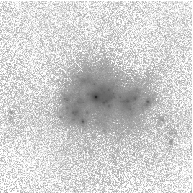}\\
\includegraphics[width=3.2cm]{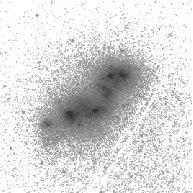}
\includegraphics[width=3.2cm]{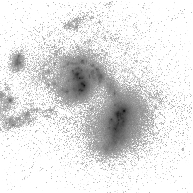}
\includegraphics[width=3.2cm]{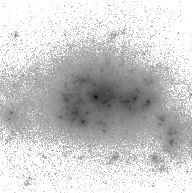}\\
\includegraphics[width=3.2cm]{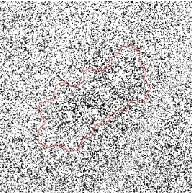}
\includegraphics[width=3.2cm]{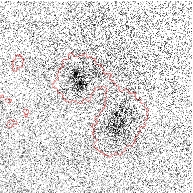}
\includegraphics[width=3.2cm]{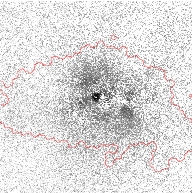}\\
\includegraphics[width=3.2cm]{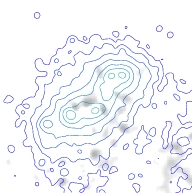}
\includegraphics[width=3.2cm]{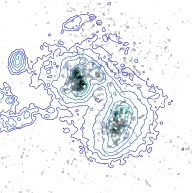}
\includegraphics[width=3.2cm]{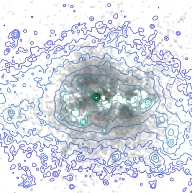}\\
\includegraphics[width=3.2cm]{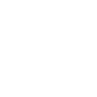}
\includegraphics[width=3.2cm]{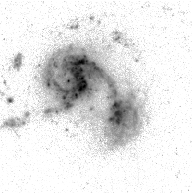}
\includegraphics[width=3.2cm]{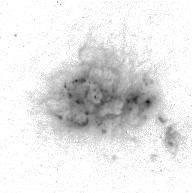}
\end{center}
\caption{Greyscale images. As in Fig.~\ref{grey1}.  
From {\em top} to {\em bottom}: F122M, F140LP, \lya, adaptively filtered \lya\ 
with F140LP isophotes overlaid, and \ha. 
From {\em left} to {\em right}: (size in arcsec in parentheses): 
\tol\ (10\arcsec) , \ngc\ (20\arcsec) and \eso\ (20\arcsec). 
The intensity scaling is logarithmic. North is up, East to the left.}
\label{grey2}
\end{figure*}

In Figs.~\ref{grey1} and \ref{grey2} we show images of our targets in the F122M 
and F140LP filters, and continuum-subtracted \lya\ and \ha\ images.
In this article we make use of  adaptive smoothing (using {\sc filter/adaptiv} 
in {\sc midas}) and adaptive binning
\citep{Diehl_Statler06}
techniques, both of which are widely used in the X-ray community. 
Smoothing results in images that are more pleasing to the eye, although less
photometrically secure. We here show \lya\ images that have been obtained by
continuum subtracting unbinned data, and an adaptively smoothed version of the same image.
F140LP FUV contours are overlaid on the smoothed \lya\ image to guide the eye.
 Binning, on the other hand, strictly conserves local surface
brightness and  has consequently been used for the photometric analysis (Sect. 
\ref{sect:res_globalphot} -- \ref{sect:localphot}, and Tables 4 and 5).

\begin{figure*}
\begin{center}
\includegraphics[width=12cm]{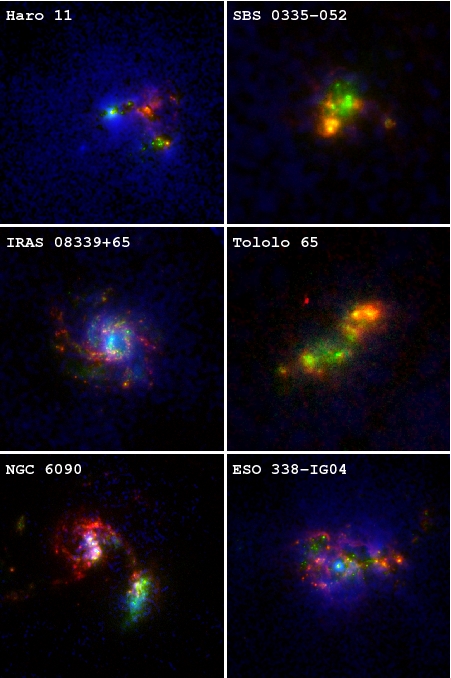}
\end{center}
\caption{Color composite images of the whole sample.
The blue channel shows continuum subtracted \lya, adaptively smoothed for the 
sake of presentation. 
Continuum subtracted \ha\ is shown in red.
1500\AA\ continuum is shown in green. 
The intensity scaling is logarithmic and arbitrary and has been adjusted to show 
interesting details. 
Each image has a size of 15\arcsec $\times$15\arcsec, except for
\sbs\ ($5\arcsec\times5\arcsec$) 
and 
\tol\ ($7.5\arcsec\times7.5\arcsec$). 
For \tol\ the red channel shows a mixture of \hb, [O{\sc iii}] 
and \ha\ (see text for details). 
East is left, North is up. 
Low surface brightness \lya\ emission can clearly be seen surrounding
compact UV point sources, particularly in \eso, \haro, and the South West knot
of \ngc. 
}
\label{rgbfig}
\end{figure*}

In Fig. \ref{rgbfig} we show RGB color composites for our six targets, 
with \ha\
in red channel, F140LP in green, and adaptively smoothed \lya\ in blue. 
The intensity scale is set to show detail and in general the \ha\ and optical emission 
can be traced to much larger radii than here apparent. 
Both greyscale and false-color composite images show \lya\ morphologies that are 
strikingly different from those in UV and \ha.  
In all galaxies we see a mixture of both emission and absorption although \lya\
appears not to emanate from regions of highest FUV surface brightness and instead
seems to be most frequently seen from regions fainter in the continuum. 

\subsection{Considerations for quantitative \lya\ photometry}\label{sect:pcyg_phot}

Absorption of the stellar continuum, both in the Milky Way and internally,
present a couple of caveats against accurate photometric analysis of the
continuum-subtracted \lya\ maps.
A review of these caveats is thus warranted.

Ultraviolet aperture spectroscopy of two of our targets (\sbs\ and \tol )
have revealed damped \lya\ absorption 
\citep{Thuan_Izotov97}. 
This is confirmed by the \lya\ continuum subtractions for  both these objects.
However, all observations in F122M are also subject to Galactic \hi\ 
absorption and  in cases where the redshift is small, internal and Milky Way 
features may be blended. 
While our continuum subtraction software does correct for the Galactic absorption, 
internal absorption features may be disguised if the features overlap. 
This is indeed the case for \tol\ and, to a much lesser extent, \sbs. 
In effect this may lead us to overestimate the emerging \lya\ flux.

In several cases when \lya\ has been detected in high resolution spectroscopy
it is found to have a P\,Cygni profile due to outflows in the neutral ISM that 
promote the escape of \lya\
\citep{Kunth98}.
When the profile is unresolved, as is the case in low-resolution spectroscopy 
or narrow-band imaging, the absorption on the blue side may cancel part of the 
emission. 
It should be noted, however, that scattering alone does not lead to a
reduction in flux, but rather a modified morphology and/or redistribution in 
frequency. 
Indeed, 
\cite{Verhamme06}
note that P\,Cygni profiles can arise as a natural consequence of radiation
transfer in a moving medium, with no net loss of photons.  

Whether moving neutral media will give rise to net emission or absorption
not only depends on the \hi\ density and outflow velocity but also on the 
relative contribution of the line and UV continuum, which of course are well
resolved at low-z. 
If the continuum is strong, as in front of UV-bright star clusters, 
then any absorption could potentially be strong as well, increasing the 
likelihood of measuring  net absorption. 
Indeed, in a few cases we find local minima in \lya\ surface brightness in 
front of UV-bright clusters that results from just this effect. 
On the other hand, in faint regions there is not much 
continuum to absorb. 
While the absorption feature itself may be undetectable in this regime, the 
blue wing of the emission profile may still be absorbed. 
When dust, of unknown distribution is added, it is impossible to know 
whether the `absorbed' photons from the blue wing in any given pixel are
(i) absorbed locally, 
(ii) scattered and absorbed elsewhere, 
(iii) scattered and emitted elsewhere, or 
(iv) scattered and redistributed in frequency to the red wing and emitted
locally. 
Note that these effects will also be significant in typical high-$z$ 
observations where the processes governing \lya\ emission must be the same.
What the true \lya\ flux really is in the case of P\,Cygni-blended absorption
and emission is a matter of definition. 
Here we naturally define \lya\ as the sum of emission and absorption.

\subsection{Quantitative global photometry and properties}\label{sect:res_globalphot}

We have extracted `global' broad-band and emission line fluxes using a 
mask which was made by cutting the 
Voronoi-binned F140LP image at a surface brightness of level of 
$\mu_{AB,1500}=23$ mag/$\square\arcsec$, corrected for 
Galactic foreground reddening. 
The global quantities presented in Tables~\ref{tab:results1} and 
\ref{tab:results2} 
have been integrated over these apertures. 
Table~\ref{tab:results1} also presents integrated far UV ($\lambda=1527$\AA) 
luminosities and UV continuum slopes ($\beta$), derived from the 
F140LP and F220W images. 
In deriving $\beta$ we used the pivot wavelengths, given in
Table~\ref{tab:results1}, but the effective wavelengths of a filter will in turn 
depend on the spectral shape, i.e.  $\beta$. Hence the  $\beta$ values
presented are only approximate, although since they are not actually used
in the analysis, this is good enough.

\begin{table*}[th]
\scriptsize
\begin{center}
\caption{Integrated broad band fluxes and magnitudes. \label{tab:results1}}
\begin{tabular}{lrrrrrrrrr}
\tableline\tableline \\
 Target & $r_{eq,1500}$ & $\lambda=1527$ &  ${2255}$ & ${3360}$  & ${4320}$ & ${5580}$ & ${8115}$ & log($L_{\rm FUV})$ & $\beta$ \\ 
        &  ($\arcsec$)             & $f_{\lambda}$ & $f_{\lambda}$ & $f_{\lambda}$  & $f_{\lambda}$ & $f_{\lambda}$ & $f_{\lambda}$ &  \\   
        &  (kpc)                   & $m_{\rm AB}$ &  $m_{\rm AB}$ & $m_{\rm AB}$  & $m_{\rm AB}$ & $m_{\rm AB}$ & $m_{\rm AB}$ &       \\  \\
\tableline \\
\haro\ & 5.4 & 5.21E-14 & 2.16E-14 & 1.09E-14 & 8.65E-15    & 5.12E-15 & 3.17E-15     & 10.3 & -2.25 \\
       & 2.2 & 14.9     & 15.0     & 14.9     & 14.6        & 14.6     & 14.3         &  & \\
\sbs\  & 3.1 & 1.24E-14 & 4.38E-15 & 1.71E-15 & 9.71E-16    & 4.08E-16 & 1.95E-16$^a$ & 9.3 & -2.66 \\
       & 0.83 & 16.4    & 16.7     & 16.9     & 16.9        & 17.3     & 17.4         &  & \\
\iras\ & 5.9 & 9.04E-14 & 3.83E-14 & 2.37E-14 & 2.50E-14    & 1.70E-14 & ---          & 10.4 & -2.21 \\
       & 2.2 & 14.3     & 14.4     & 14.0     & 13.4        & 13.3     & & & \\
\tol\  & 2.3 & 4.26E-15 & 1.91E-15 & 7.46E-16 & 4.65E-16    & 1.95E-16 & 9.30E-17$^b$ & 8.5  & -2.05 \\
       & 0.42 & 17.6    & 17.6     & 17.8     & 17.7        & 18.1     & 18.1         &  & \\ 
\ngc\  & 4.0 & 1.71E-14 & 9.86E-15 & 6.81E-15 & 6.84E-15    & 5.18E-15 & 3.60E-15~    & 10.1 & -1.41 \\
       & 2.3 & 16.1     & 15.8     & 15.4     & 14.8        & 14.6     & 14.2         &  & \\
\eso\  & 7.6 & 1.11E-13 & ---  & 2.16E-14 & 1.81E-14$^c$ & 1.25E-14 & 5.76E-15$^b$ & 9.9  & -2.07 \\
       & 1.4 & 14.1     &       & 14.1     & 13.7        & 13.6     & 13.6        \\ \\
\tableline
\end{tabular}
\tablenotetext{a}{Wavelength 7880}
\tablenotetext{b}{Wavelength 8012}
\tablenotetext{c}{Wavelength 4310}
\tablecomments{Integrated broad band fluxes and magnitudes within the isophotal
mask ($\mu_{\rm 1500, AB}=23$ mag/arcsec$^2$). 
The second column gives the equivalent radius ($r_{eq}=\sqrt{{\rm Area/\pi}}$) within which the fluxes 
were integrated in units of arcseconds (first line) and kpc (second line).
First line gives $f_\lambda$ in cgs units (erg~s$^{-1}$~cm$^{-2}$~\AA$^{-1}$), 
second line gives magnitudes in the AB-system. $L_{\rm FUV} = \lambda\times L_\lambda$ 
in units of $L_{\odot, \rm Bol} = 3.8\cdot10^{33}$ erg/s, and $\lambda=1527$\AA. 
The UV-continuum slope is derived from $f_{1500}$ {\em vs.} $f_{2200}$
except for ESO\,338--04 where we used $f_{3360}$ rather than the available F218W image which 
is not as deep and quite affected by CR-residuals. All quantities have been corrected for 
Galactic extinction using the Schlegel et al. (1998) prescription.}
\end{center}
\end{table*}

\begin{table*}[th]
\scriptsize
\begin{center}
\caption{Measured emission line properties\label{tab:results2}.   }
\begin{tabular}{lrrrrrrrrrr}
\tableline\tableline  \\ 
Target name & 
$f_{\mathrm{Ly}\alpha}$ &  
$W_{\mathrm{Ly}\alpha}$ & 
$L_{\mathrm{Ly}\alpha}$ & 
$\sigma_{\mathrm{Ly}\alpha}$ & 
$f_{\mathrm{H}\alpha}$ & 
$W_{\mathrm{H}\alpha}$ & 
$L_{\mathrm{H}\alpha}$ & 
${\mathrm{Ly}\alpha}/{\mathrm{H}\alpha}$ \\ \\
\tableline  \\
\haro\  & 9.30E-13   & 15.6   & 8.40E+41   & 4.0E-15 & 2.48E-12 & 704  & 2.24E+42 & 0.374    \\
    {\it ~+ only}    & $\mathbi{4.32E-12}$   & $\mathbi{25.5}$   & $\mathbi{3.90E+42}$   &      &       &  &       & $\mathbi{0.625}$   \\ \\ 
\sbs\   & --6.77E-13 & --36.8 & --2.59E+41 & 2.9E-15 & 3.61E-13 & 1434 & 1.38E+41 & --1.76   \\
    {\it~+  only}   & $\mathbi{2.97E-14}$   & $\mathbi{1.62}$   & $\mathbi{1.12E+40}$   &      &       &  &       & $\mathbi{0.078}$   \\  \\ 
\iras\  & 3.05E-12   & 39.2   & 2.37E+42   & 3.4E-15 & 2.37E-12 & 199  & 1.84E+42 & 1.29     \\
    {\it ~+ only}    & $\mathbi{3.52E-12}$   & $\mathbi{40.7}$   & $\mathbi{2.73E+42}$   &      &       &  &       & $\mathbi{1.54}$     \\ \\ 
\tol\   & 8.48E-14   & 17.5   & 2.82E+40   & 2.3E-15 & 1.85E-13 & 1300 & 3.41E+40 & 0.458    \\
    {\it~+  only}   & $\mathbi{1.47E-13}$   & $\mathbi{15.5}$   & $\mathbi{4.89E+40}$   &      &       &  &       & $\mathbi{0.794}$    \\ \\ 
\ngc\   & 4.20E-13   & 28.3   & 7.80E+41   & 1.7E-15 & 1.27E-12 & 315  & 2.36E+42 & 0.330    \\
    {\it~+  only}   & $\mathbi{5.71E-13}$   & $\mathbi{38.2}$   & $\mathbi{1.06E+42}$   &      &       &  &       & $\mathbi{0.449}$    \\ \\ 
\eso\   & 3.68E-12   & 28.1   & 6.89E+41   & 3.7E-15 & 2.61E-12 & 570  & 4.89E+41 & 1.41     \\
    {\it~+  only}   & $\mathbi{4.99E-12}$   & $\mathbi{38.0}$   & $\mathbi{9.35E+4}$1   &      &       &  &       & $\mathbi{1.91}$     \\  \\ 
\tableline
\end{tabular}
\tablecomments{Integrated galaxy properties within the far UV isophotal mask 
($\mu_{1500, AB}=23$ mag/$\square\arcsec^2$). 
$\sigma_{Ly\alpha}$ is the `spaxel--to--spaxel' standard 
deviation of the \lya\ background outside the photometric aperture and hence 
measures the fluctuations on a one arcsecond scale. 
Note that this does not entirely reflect the error from flatfielding and
sky subtraction which is conservatively found  to be smaller than 20\%.
All quantities have been corrected for Galactic extinction using the 
Schlegel et al. (1998) prescription. 
Fluxes and luminosities are in cgs units, and equivalent widths in \AA.
The \ha\ luminosity and equivalent width of \tol\ has been derived from 
ground based observations (see text).
The second line for each galaxy gives the result when including pixels with 
positive \lya\ flux only.}
\end{center}
\end{table*} 

Table~\ref{tab:results2} presents the derived \lya\ and \ha\ quantities.
Regarding \lya, we firstly quote the measured continuum-subtracted flux, 
\flya, which is the sum of all pixels (positive and negative) within the 
aperture.
In addition we quote the sum of only the positive pixels, \flyapos, thereby
rejecting all pixels that represent absorption of the stellar continuum. 
Note that our ability to resolve 
emission and absorption on small scales depends on the distance to the target. 
The integrated \lya\ flux within our apertures is in five cases positive and 
one case negative.
The \lya/\ha\ ratios are an order of magnitude lower than the recombination
value of 8.7 (Case B,
\citealt{Brocklehurst71}).
This indicates significant attenuation of \lya\ due to dust, possibly aided by
radiative transfer effects from \hi.
The \lya\ equivalent widths ($W_{\lyam}$) range
from --37 to +39 \AA , i.e. significantly smaller than for the theoretical
predictions for young starbursts. To first order equivalent widths should
be unaffected by reddening if the line and continuum are equally attenuated.
The \ha\ equivalent widths ($W_{\ham}$) appear anti-correlated with $W_{\lyam}$ if
anything (see Fig \ref{fig:globals}).
For pure recombination 
one would expect $W_{\ham}$ and $W_{\lyam}$ to be positively correlated, so
clearly other effects than ordinary dust reddening must be at work.

In Table~\ref{tab:results3} we present quantities directly related to the
escaping fractions of \lya\ and dust extinction, integrated over the whole 
region of the mask. 
Firstly we present the \lya\ escape fraction which has been calculated from:
\begin{equation}
f_{esc,\lyam} = \sum f_{\lyam} / (8.7\times \sum f_{\ham,C}) 
\end{equation}
where $f_{\ham,C}$ is the \ha\ flux corrected for the internal reddening found
in our SED fit. 
This reddening is actually that of the stellar continuum and may hence be an 
underestimate for the nebular gas \citep[e.g.][]{Calzetti00}, which may in turn
cause us to overestimate $f_{esc,\lyam}$. 
See 
\cite{Atek08} 
for an analysis of the nebular gas extinction in these 
galaxies. We find that $f_{esc,\lyam}$  varies between 0 (\sbs; actually absorption so 
the measured escape fraction is negative) and around 14\% (\iras and \eso).
$f_{esc,\lyam}^+$ is also presented, where only positive \lya\ pixels are 
included in the summation.
Rejection of negative pixels results in little change in the \lya\ escape fraction 
for the galaxies that show few or weak absorption centers, (e.g. \iras), significantly 
increases it in some cases (e.g. \haro\ and \tol), and changes its sign in the
case of \sbs, for which the escaping fraction of detectable photons is around 2\%.
In no cases does $f_{esc,\lyam}^+$  reach above 20\%.

Since we have an estimate of the internal reddening, we deem it meaningful to
also use this value to de-redden the \lya\ fluxes. 
 This makes it possible to compute dust-corrected \lya /\ha \ ratios and 
dust-corrected \lya\ escape fractions.  Of course, if standard 
recombination theory was directly applicable, the corrected \lya /\ha \ ratio should 
be 8.7 and the dust-corrected \lya\ escape 
fractions should be unity.   
Instead, the extinction-corrected \lya\ fluxes are well below standard recombination
predictions with factors ranging from 2.6 to infinity. In Table \ref{tab:results3} we
present $\mathcal{R} =  \sum f_{\lyam,C} / (8.7\times \sum f_{\ham,C})$, i.e. the ratio 
of the extinction-corrected  \lya /\ha \  to the recombination value and 
equivalently  refers to the extinction-corrected `escape fraction'

Comparing $f_{esc,\lyam}$ with $\mathcal{R}$  reveals that (for galaxies
with net emission) the latter is a factor 2 to 5 larger than the former.
 \eso\ shows a \lya\ escape fraction of  almost 40\% after de-reddening,
and \iras\ approaches 30\%. The ratio of $\mathcal{R}$ to $f_{esc,\lyam}$ 
is not significantly different when regarding positive pixels only,
and therefore not presented.
The emerging picture is hence that simple reddening corrections can't come 
close to explaining the found escape fractions, meaning that in addition to
simple dust reddening, other effects, i.e. resonant 
scattering and ISM kinematics, appear to dominate the \lya\ transport.

Of course, these \lya\ escape fractions would be even lower if we have underestimated 
the total  ionizing continuum radiation, i.e. if the true reddening of 
\ha\ photons is larger than found by our SED fitting which could happen if
there were some very deeply embedded sources that are completely absorbed 
in the UV. In \cite{Atek08}  we use ground based \ha\ and \hb\ images to
estimate the dust extinction and consistently find even smaller escape fractions.

\begin{table*}[th]
\scriptsize
\begin{center}
\caption{Derived and reddening corrected properties\label{tab:results3}. }
\begin{tabular}{lrrrrrrrrrr}
\tableline\tableline \\
Target & $f_{esc,Ly\alpha} (+)$ & $\mathcal{R}$ & SFR$_{H\alpha}$ & log($L_{\rm FIR}$) & SFR$_{H\alpha}$/SFR$_{\rm FIR}$ & $E(B-V)_\star$ & log(L$_{\rm FUV,C}$)  \\ \\
 \tableline  \\

\haro\  &  0.037 (0.062)    & 0.17    & 17.2 & 11.1 & 1.3~~~~~~ & 0.067 & 10.6 \\
\sbs\   &  --0.203 (0.009) & --0.28  & 1.1  & 9.3 & 2~~~~~~   & 0.044 & 9.45  \\
\iras\  &  0.140 (0.168)     & 0.28     & 14.1 & 11.0 & 0.9~~~~~~ & 0.034 & 10.6 \\
\tol\   &  0.053 (0.091)   & 0.16     & 0.26 & 8.3 & 5~~~~~~   & 0.11 & 8.94  \\
\ngc\   &  0.027 (0.037)   & 0.13     & 18.2 & 11.4 & 0.4~~~~~~ & 0.19 & 10.8 \\
\eso\   &  0.139  (0.189)     & 0.39     & 3.9  & 9.8  & 3.4~~~~~~ & 0.076 & 10.2 \\  \\
\tableline
\end{tabular}
\tablecomments{$f_{esc,Ly\alpha}$ is the \lya\ escape fraction obtained by 
dividing the observed (Galactic extinction-corrected only) \lya\ flux by the 
\ha\ flux corrected for Galactic {\em and internal} reddening (as a measure 
of the total intrinsic \lya\ production) and the theoretical line ratio.  
$\mathcal{R}$ is the same quantity but using the \lya\ flux corrected 
for internal extinction. 
SFR$_{H\alpha}$ is based on the observed \ha\ flux in Table 
\ref{tab:results2} (not corrected for internal reddening) and uses the 
calibration by Kennicutt (1998).
log$(L_{FIR})$ is based on the 1.7 times the luminosity at 60$\mu$m 
(Chapman et al. 2000) 
as detected by IRAS and listed by NED, except for \sbs\ and \tol\ which
were not detected by IRAS and where we have instead used the Spitzer values reported by 
Engelbracht et al. (2008). 
The FIR SFR calibration in SFR$_{H\alpha}$/SFR$_{\rm FIR}$ uses the calibration
by Kennicutt (1998). 
$E(B-V)_\star$ is the  color excess found in the SED fit for the integrated
fluxes within the same far UV isophote as above.}
\end{center}
\end{table*} 

\subsection{Small scale photometry and properties }\label{sect:localphot}

\begin{figure*}
\begin{center}
\includegraphics[width=12cm]{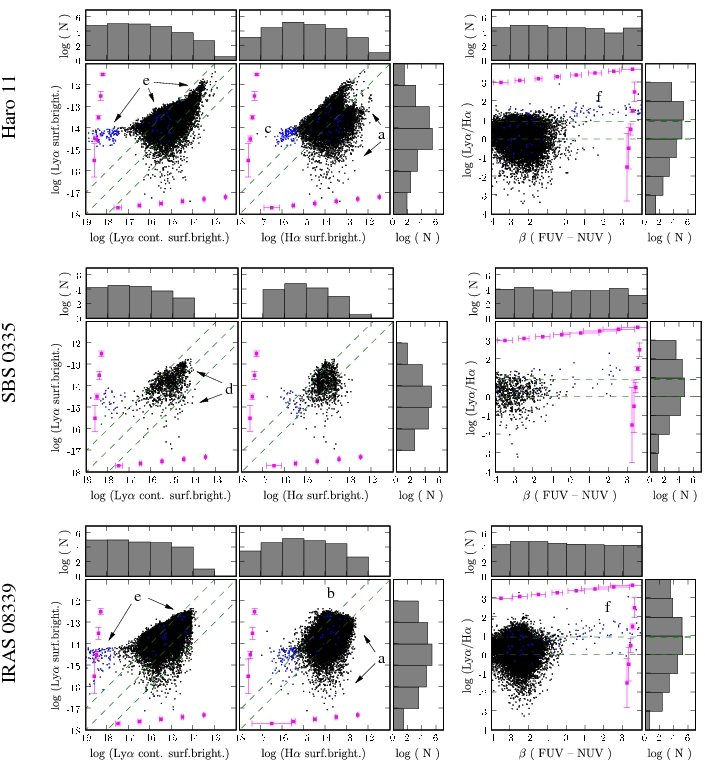}
\end{center}
\caption{Pixel scatter distribution showing all `spaxels' for which $S/N\ge3$ in F140LP.
Each spaxel, which may contain up to 1600 pixels of size 0.025\arcsec\ is
represented by a single point.
Black points show spaxels where $S/N\ge 10$ was attained, blue points show pixels in the range $3 < S/N < 10$. 
For the histograms, however, each spaxel is represented by the number of pixels 
that contributed to it, in order to show the true distribution. 
Since the ordinate shows the logarithm of the \lya\ surface 
brightness, only spaxels with positive \lya\ flux are included. 
Error-bars show the median statistical error in each magnitude bin for the ordinates and abscissae
(placed diagonally to avoid overlap). 
{\em Left:} \lya\ {\em vs.} UV surface brightness, here represented by the modeled 
continuum at 1216\AA , i.e. $\mu_{\lambda 1527}\times CTN$.
The dashed lines show constant \lya\ equivalent widths of 1, 10, and 100 \AA.
{\em Center:} \lya\ {\em vs.} \ha\ surface brightness, both in units of erg/s/cm$^2$/arcsec$^2$. 
The dashed lines show $\lyam = \ham$, and  $\lyam = 8.7\ham$ (Case B recombination).
{\em Right:} \lya/\ha\ line ratio {\em vs.} UV continuum slope $\beta$ as derived from the fluxes in 
F140LP and F220M. 
{\em Top panels:}    \haro. 
{\em Middle panels:} \sbs.
{\em Bottom panels:} \iras.
\label{sc1} }
\end{figure*}

\begin{figure*}
\begin{center}
\includegraphics[width=12cm]{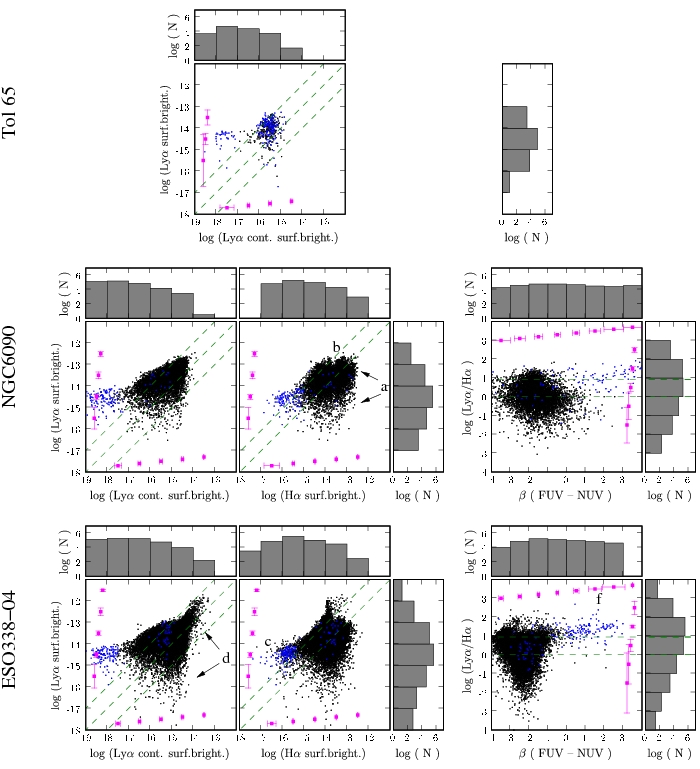}
\end{center}
\caption{Same as Fig. \ref{sc1} but for \tol\ (top), \ngc\ (middle) and \eso\ (bottom).
For \tol\ we do not show \ha\ since we only have reliable data at ground based resolution.
For \eso\ $\beta$ was derived from the F140LP and F336W data. 
See caption of Fig. \ref{sc1} for further details.
\label{sc2} }
\end{figure*}

In Figures~\ref{sc1} and \ref{sc2} we show the \lya\ surface brightness
{\em vs.} \ha, UV continuum, and $\beta$. 
The Voronoi binned resolution elements used to measure surface brightness vary
in size between individual pixels in the central regions and 1600 pixel
(1$\square\arcsec$) spaxels in the faintest regions. 
Here we do not mask out the fainter regions as described for the integrated
photometric measurements. 
Instead, for inclusion of a spaxel, the criterion of having reached $S/N\ge 3$ in 
the F140LP image must have been fulfilled. 
This allows us to probe the fainter surface brightness regions while 
maintaining data reliability. 
 Resolution elements in which the desired threshold $S/N\ge 10$ 
\citep[See][]{Hayes09}	
has been met are shown in black, whereas pixels with $3 \le S/N < 10$ are shown in 
blue.
Since the pixel distribution plots use bins of various sizes, they cannot be
used to estimate how the values are globally distributed and instead, histograms are
presented above and to the side of the scatter plots in order to give some
feeling for this. 
Moreover, since the surface brightnesses are presented in log-space, only pixels
showing emission are included. 
Dashed lines on the \lya\ {\em vs.} \ha\ plots show the recombination line
ratio of $f_{\rm Ly\alpha}=8.7\times f_{\rm H\alpha}$ and also the  case
$f_{\rm Ly\alpha}= f_{\rm H\alpha}$. 
On the \lya\ {\em vs.} FUV plot the lines represent constant equivalent
width of 1, 10, and 100\AA. 
These scatter plots show a number of characteristic features which we now
discuss. 
For illustration several features are labeled (a--f) in the text and in 
Figures~\ref{sc1} and \ref{sc2}.  

\paragraph{\lya\ {\em vs.} \ha :}
A general feature is a sharp boundary at maximum \ha \ surface brightness, 
at which \lya / \ha\ ranges from unity and downwards (a), except for \sbs\ where
\lya / \ha\ reaches the recombination value. The large range of values (extending
to negative values, but which are not shown in these log-plots)  is either the
 result of dust reddening or resonant scattering decoupling the \lya\ photons from \ha.
Another general feature is a ceiling in \lya\ surface brightness at 
$\sim 10^{-13}$ erg/s/cm2/arcsec$^2$ (b). 
While  \haro\ and \eso\ present similar features, we see, in
addition, a population of \lya -bright spaxels with \lya / \ha\ at, or above,
the recombination value (c). In all galaxies an upper envelope is seen near or slightly
above the recombination ratio, but at fainter \ha\ levels a tail with super-recombination
values develops -- an indication of resonant scattering.

\paragraph{\lya\ {\em vs.} UV continuum:}
A rather sharp boundary is seen at UV-bright levels ($\sim 10^{-14}$ to $10^{-15}$ 
erg/s/cm$^2$/$\square\arcsec$/\AA ), reminiscent 
of a starburst intensity limit (d). From the combined UV and \lya\ peak towards 
fainter UV surface brightness levels we see an upper envelope, which starts
at $W_{\lyam}\sim30$\AA\ at the bright end and approaches 1000\AA\ at fainter
($\sim 10^{-17}$erg/s/cm$^2$/$\square\arcsec$/\AA ) levels where it forms a tail 
with $W_{\lyam} > 1000$\AA\ (e). 
   Again \haro\ and \eso\ present extra high surface 
brightness features with average equivalent widths of about 30 \AA.

Hence, we see regions where line ratios and equivalent widths are significantly enhanced 
compared to what would be expected from recombination theory and standard spectral synthesis
models. This may naturally be expected to some extent for \lya\ {\em vs.} FUV since
ionized 
regions may be spatially separated from the UV, but for \lya\ {\em vs.} \ha\ it must imply
that some process, resonant scattering, is causing \lya\ to be emitted from
regions where \ha\ is very faint indeed.

\paragraph{\lya / \ha\ {\em vs.} ${\boldmath \beta}$:}
Pixel--to--pixel variations between the \lya/\ha\ ratio and $\beta$ again show
little correlation as if \lya\ emission was independent of the local stellar reddening.
A tail in the distribution is also seen in these figures 
up to $\beta > 0$ (f). 
These red UV slopes indicate that either the stellar population is too old to
produce nebular emission, or the dust reddening is too great to permit the
transmission of \lya. 
Either way, \lya\ is seen in emission from these regions, and a significant
fraction of the pixels show ratios greater than that of case B recombination.

\begin{figure*}
\begin{center}
\includegraphics[width=15cm]{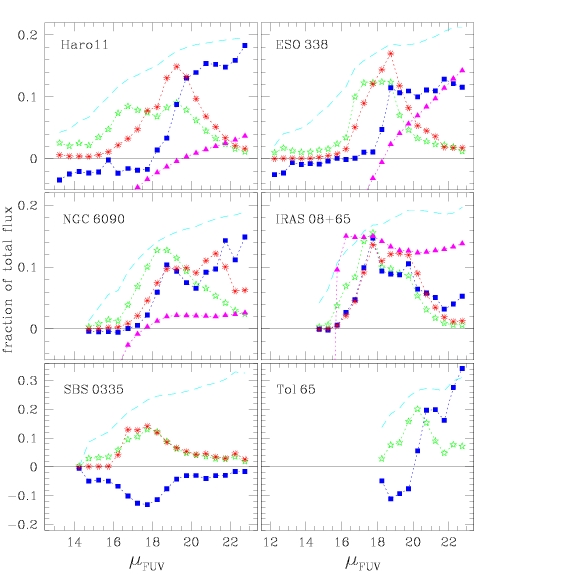}
\end{center}
\caption{Flux distribution vs far UV surface brightness.
Here we show the fraction of the total (within the $\mu_{AB,1500}=23$ mag$/\square\arcsec$ mask) 
flux  as a function of $\mu_{AB,1500}$ (corrected for Galactic reddening only) summed
in 0.5 magnitude bins.
Stars (green) show 1500\AA, filled squares (blue) show \lya, and asterisks (red) show \ha.
Triangles (magenta) show the cumulative \lya\ escape fraction, determined from
\ha\ -- they always converge at the escape fraction presented in
Table~\ref{tab:results3}. 
Since the aperture mask is defined by making a surface brightness cut in F140LP,
this demonstrates how the \lya\ escape fraction may vary had this cut-off been chosen
differently. 
The dashed lines (cyan) show $\log(N_{\rm pix})/C$, the logarithm of the number of pixels in 
each 0.5 magnitude which, for the purpose of presentation, has been divided by a number $C$ 
($C=22.5$ for all galaxies but \sbs\ and \tol\ where it is 12.5). 
}
\label{fig:fluxdist}
\end{figure*}

\paragraph{UV surface brightness:}
In Fig. \ref{fig:fluxdist} we show the fraction of \lya\ fluxes emitted (or absorbed) 
within the masked regions as a function of the UV surface brightness.
This is done by summing the flux in the pixels of the line-only image that fall
within the surface brightness bins in the FUV image, then normalizing by the
integrated flux within the masked region.
Thus we can quantitatively assess how \lya\ emission/absorption correlate with 
UV surface brightness.
The same is also presented for \ha\ and UV continuum itself.
In most cases \ha\ is clearly seen to be offset from the UV continuum since, to
some varying degree, the bright stellar sources and ionized nebulae can be resolved. 
While the bulk of the emitted \lya\ flux also appears to emerge from regions of lower 
FUV surface brightness, the distribution is further offset from that of \ha.

\subsection{Comments on individual targets}\label{individuals}

\paragraph{\haro}
This target is extensively discussed in 
\cite{Hayes07}, 
where 
we also discuss X-ray data, the escape of \lya\ as a function of the age of 
the stellar population (field stars and stellar clusters) and nebular reddening (from 
ground based observations). 
\cite{Kunth98}
found an asymmetric \lya\ line in emission with no sign of \lya\ emission at restframe 
1216~\AA, suggesting \lya\ escape due to an outflowing neutral medium. 
In addition blueshifted metallic absorption lines indicated the presence of 
outflows along the line of sight. 
\cite{Bergvall06} claimed a detection of Lyman continuum (LyC) photons escaping from this 
galaxy (but see also Grimes et al. 2007) and, since LyC does not scatter, porous 
ionized holes in the ISM are likely 
to exist along some sight-lines. 
The galaxy is very luminous in \ha\ and has a high inferred star 
formation rate, still no H{\sc i} or CO has been detected 
\citep{Bergvall03}, 
the latter may be related to the low metallicity 
\cite{Bergvall_Ostlin02}.

This galaxy consists of three major star forming condensations although \lya\ is
seen directly in emission from only one (the NW knot; C in the nomenclature of
\citealt{Kunth03}).
It is this knot that causes the high surface brightness features discussed 
in Section~\ref{sect:localphot}.
The remaining two knots (one of which dominates the \ha\ output) show \lya\ absorption 
centered on the clusters themselves, with a patch of faint diffuse emission leaking from 
the condensation to the SW. 
The overall \lya\ flux escaping is dominated by a diffuse 
component with \lya/\ha\ ratios larger than the recombination value, 
indicating that most \lya\ leaves the galaxy after multiple scatterings (see 
\cite{Hayes07} 
for a description of the \lya\ surface brightness profile). 
In the \lya-bright eastern knot \lya/\ha\ is close to the recombination
value, and the EW is high, indicating that \lya\ emission from this region may
be avoiding resonant trapping and escaping more or less unhindered.

\paragraph{\sbs} 
Previous attempts have failed to detect anything 
but damped \lya\ absorption from this galaxy 
\citep{Thuan_Izotov97}. 
Here we show, that while the central region is indeed characterized by \lya\ 
absorption there are regions from which \lya\ photons are emitted. Compared to the
other galaxies, the fraction of \lya\ emitting spaxels is small, but significant
at  the $3 \sigma$ level.
 Integrated over the galaxy, it is a net absorber of \lya.

The brightest \lya\ emitting spaxels have \lya/\ha\  somewhat higher 
than the recombination value, which could result from resonant scattering, ISM 
geometry, or the production of additional \lya\ photons by collisional excitation.  
This galaxy exhibits a low surface brightness tail just as the brighter \lya\
emitters do. 

The lack of a net positive escape at any surface brightness level is somewhat
puzzling as the \lya\ photons ought to get out somewhere, unless the density of 
static H{\sc i} is so large that even moderate amounts of dust will destroy
\lya.

\paragraph{\iras} 
This is a spiral with a nuclear starburst. 
\cite{Kunth98} 
found P-Cygni like emission, and 
\cite{Mas-Hesse03}
found further support for an ISM-kinematic regulation of the \lya\ escape and
evidence for recombinations in the outflowing shell.
\lya/\ha\ again supports a mixture of dust and kinematic effects in the ISM.

Since no F814W image was obtained for \iras, our data-points are not
sufficiently spread along the SED to fit two stellar populations, and continuum
subtraction of \ha\ also becomes less reliable. 
For \iras\ we therefore resort to a single component spectral fit using the
standard stellar+nebular template from {\em Starburst\,99}, i.e. method {\sc iii}
in the terminology of Paper\,I.

\paragraph{\tol} 
This is the faintest galaxy in the sample, both in luminosity and apparent magnitude. 
In our HST \ha\ image, the galaxy happened to fall outside the monochromatic patch of the ramp 
filter and could not be used. Hence, we had no information about the distribution of ionized
gas at HST resolution and could not model the nebular emission as a separate component. Instead
we used two stellar components where we used the option of {\em Starburst\,99} to include the 
nebular continuum  in the model \citep{Leitherer99}. In addition, the ACS/SBC observations of
\tol\ suffered a guide star acquisition failure meaning that only one of the two guide stars 
were acquired, with the consequence that the telescope pointing could be drifting about the axis 
defined by the direction to  the single guide star used. The images were registered manually
rather than using {\sc multidrizzle} since the astrometric keywords could not be trusted.
We have carefully examined the SBC images and found no evidence for a significant telescope
drift during the observations, but the results for \tol\ should be viewed with some caution.

To get a rough idea about the ionized gas distribution we retrieved a WFPC2 F555W image 
of exposure time 1600s from the archive. 
The F555W and F550M filters have very similar central wavelength but,  
whereas F550M does not transmit any strong emission lines, F555W transmits 
\hb, [O{\sc iii}]$\lambda\lambda 4959,5007$, and partly  \ha. 
Thus, by subtracting F555W--F550M, scaled by their PHOTFLAM values,  we get a map 
of the strong emission lines, that can be used for qualitative purposes, 
e.g. in making the color image in 
Fig.\ref{rgbfig}. 
For quantitative estimates of the \ha\ flux and equivalent width, we have used 
ground-based data obtained from ESO's New Technology Telescope (NTT)
\citep{Atek08}.  
Previous HST/GHRS spectra 
\citep{Thuan_Izotov97} 
have indicated that the brightest
knot is a damped \lya\ absorber. The UV morphology of Tol\,65 is also rather 
complex but, assuming that the NW knot was acquired by the GHRS PEAK-UP
maneuver, photometry performed in a $2\arcsec\times2\arcsec$ aperture consistently 
finds this to be an absorber.

Overall we find that the continuum-bright regions are dominated by absorption but
we also see \lya\ in emission from fainter regions. \tol\ is the galaxy for which the data quality is worst and the uncertainties 
are largest.
Due to the effects discussed in Sect.~\ref{sect:pcyg_phot}, the positive net \lya\ flux observed from this galaxy may not be real.

\paragraph{\ngc}
This luminous starburst has a clear merger signature and
is also very luminous in the IR. 
The star formation is concentrated to two regions both which show clear \lya\ emission
with emission/absorption varying on small scales.
The \lya/\ha\ ratio scatters between above the recombination value and absorption.
The common tail of \lya/\ha\ ratios is seen, extending to values that exceed 
the recombination value, and representing diffuse regions outside the star forming complexes.

\paragraph{\eso}
The \lya\ morphology of this galaxy was discussed in
\cite{Hayes05} using a less sophisticated method 
(similar to method {\sc iii} in the terminology of Paper\,I) for continuum subtraction
which was based on a single stellar population and
 did not include a separate treatment of the nebular continuum.
Our new results are, however, in good agreement with 
\cite{Hayes05}.
The starburst
is dominated by a rich population of young massive star clusters 
\citep{Ostlin98,Ostlin03}, 
around which strong \ha\ is seen following a rich and detailed structure
consisting of shells and filaments. 
In contrast the \lya\ emission is much more extended and clearly 
asymmetric with the main emission in the N--S direction, where there are signs of 
outflows from other investigations 
\citep[e.g.][]{Ostlin01,Ostlin07}.
Central patches of \lya\ absorption are seen across much of the central star
forming complex with a noticeable streak in the E--W direction. 
This is not seen in age or \ebv\ maps and is most likely the result of a
filamentary \hi\ structure lying in front of the starburst. 
Unique for this galaxy is the clear vertical tail to super recombination values 
in the \lya vs \ha\ distribution occurring also at bright levels (Knot A in the 
nomenclature of \cite{Hayes05}. See Fig. \ref{sc2}).

\section{Discussion}\label{sect:discussion}

All targets reveal \lya\ morphologies qualitatively very different from those
revealed by the FUV continuum or \ha.
Two \lya\ non-emitters (\sbs\ and \tol) were on purpose included in the sample, 
but both show signatures of \lya\ leakage, and one of them is actually found to
be a net \lya\ emitter (\tol), although this is our object with the poorest data quality.
In the remaining cases (\haro, \iras, \ngc, \eso) where the \lya\ signature is 
very clear and well-structured, the equivalent widths are found to be low with 
line-ratios significantly lower than the recombination value, and
escape fractions well below unity. 
		
Our targets span luminosities from the dwarf regime to that of $z\sim 0.15$ compact UV-luminous 
galaxies (UVLGs) that have been proposed as local analogs of LBGs by virtue of their
luminosity and morphology 
\citep{Overzier08}. 
Indeed,  \haro\ was used in their 
paper as a local comparison object. Although their selection did not favour inclusion
of galaxies with multiple nuclei, such as \haro\ or \ngc\, our four most luminous galaxies
otherwise have very similar properties to UVLGs. Morphologically, and in terms of 
rest rest-frame colors and luminosities our sample galaxies are also similar to 
starbursts at $z\sim 1$ studied in deep HST U-band images 
\citep{deMello06}. 

\subsection{Emission scales and porosity } 

In front of the star forming regions we find strong variations in both the \lya\ 
images and equivalent width maps, ranging between absorption and emission. 
This phenomenon is seen on scales ranging between the smallest we can probe
reliably (0.1\arcsec, corresponding to a physical sizes ranging between 20 and 60 pc) 
up to sizes an order of magnitude larger. 
The scatter plots (Figs. \ref{sc1} and \ref{sc2}) clearly show the large variation 
of \lya\ equivalent width (for positive spaxels) from the regions of peak UV and \ha\ 
surface brightness.
There is no apparent characteristic size over which these variations are seen.
That is, the \lya\ flux or $W_{\lyam}$ may be radically different, 
e.g. switching from absorption to emission, even on the smallest scales we can 
probe.  
This supports the model in which \lya\ may escape through favorable paths in a
porous and inhomogeneous ISM.
However, it is also clear that this is not the main mechanism by which 
\lya\ photons escape galaxies, because the net emission, if any, is always
dominated by diffuse emission from regions with low UV and \lya\ surface brightness.

In Figure~\ref{fig:fluxdist} we have presented the fractional contribution to the
total \lya\ and \ha\ emission as a function  of UV surface brightness.
Four out of five targets  (\tol\ was not observed in \ha ) show a systematic 
offset between \ha\ and FUV (red lines appear 
shifted towards fainter levels
compared to green), which can be expected for two reasons: 
Firstly, the ionization boundary will be extended away from  
clusters, where UV surface brightness is lower and may be resolved in the images.
Secondly, stellar mechanical feedback may blow out the ionized ISM in shells
and filaments, further decoupling the nebular and stellar components.
Likely this is true also for \tol\ since, like in the other four cases, 
\lya\ is offset with respect to FUV. 
\lya\ is in general further (compared to \ha ) offset towards lower FUV surface
brightness levels, demonstrating the importance of additional scatterings 
affecting \lya. 

The only object that does not show a systematic offset between FUV and \ha\ surface 
brightness is \sbs.
\cite{Mas-Hesse03}
suggested that damped \lya\ absorption profile observed in this galaxy to be due
to the youth of the starburst episode ($W_{\ham}=1500$\AA); 
insufficient time has elapsed to generate a strong wind and accelerate the 
surrounding \hi.
The fact that we observe \ha\ and FUV emission from \sbs\ to be rather tightly 
correlated could also  be a reflection of this fact, and is consistent with the 
interpretation.
The \lya\ line, on the other hand, almost completely mirrors that of the FUV, and 
is seen only in strong absorption in the central regions. 
This is naturally an effect of absorption of the stellar continuum as discussed in 
Section~\ref{sect:pcyg_phot}, where more flux appears to be absorbed at higher surface
brightness.

\iras\ and, to a lesser degree, \ngc\ 
shows some degree of correlation between the relative \ha\ emission and \lya. 
However, at intermediate UV surface brightness ($\mu_{AB,1500}=18$ to $20$ mag/$\square\arcsec$) 
systematically less \lya\ is emerging with respect to \ha\ than at fainter continuum levels. 
Again this is indicative of the resonant scattering of \lya\ increasing typical
physical 
scales of the emission and resulting in low-level emission outside the starburst 
region. 

The remaining strongly emitting galaxies, \haro,  and \eso\ show strong and
systematic offsets between the surface brightnesses where the bulk of the \ha\
and \lya\ emerge. 
This is nicely illustrated also in Figures~\ref{grey1},\ref{grey2}, and \ref{rgbfig}
where large scale \lya\ emission is seen from outside the star forming regions.
In \haro, the peak in the output \lya\ luminosity distribution is offset from
\ha\ by about 2.5 magnitudes in $\mu_{AB,1500}$. 
In \eso\ the bulk of the \lya\ output is shown clearly to be at faint UV flux
levels although no clear maximum is seen and much \lya\ emerges from the faintest
UV bin. 

Clear signatures of diffuse \lya\ emission, above the level of the background
residuals is seen in both \haro\ and \eso\ at radial distances from the central
star-forming region of around 3 and 1.5~kpc, respectively.

\subsection{\lya\ escape fraction and dust}

Unlike non-resonant radiation, photon scattering causes \lya\ to be
emitted over larger areas than radiation in the adjacent continuum and at lower
local signal--to--noise where it is more susceptible to contamination by
residuals from the sky-background and flatfielding.
Thus the measured \lya\ escape fraction may be a strong function of aperture 
which is evident from Fig
\ref{fig:fluxdist}.
Had our data been deeper and we could have been more confident about photometry
at very faint levels, the escape fractions may well have increased. 
For a case like \sbs, which is not very dusty, one would naively expect most of the 
\lya\ to emerge somewhere -- indeed a significant fraction of \ha\ emission comes 
from outside our aperture mask.
On the other hand, when we fit a S{\'e}rsic profile to the \lya\ scattering
component of \haro\ at good S/N
\citep{Hayes07}
and integrated the flux to an infinite radius, we obtained a flux consistent
with aperture photometry within a few per-cent.  
For reference, to reach S/N=10 in a $1 \square\arcsec$ aperture, a source 
of the same surface brightness as the average spaxel--to--spaxel sky variation
(see Table~\ref{tab:results2}) would require over 30 orbits in SHADOW mode with
SBC/F122M per target.  

Five out of six cases are found to be net \lya\ emitters with \lya\ escape fractions in the 
range 3 to 14 percent. 
From the sixth galaxy (\sbs), we do detect some  \lya\ emission leaking outside
the central starburst region.
However, also at faint levels is the net 
integrated \lya\ flux negative.
It is possible that more \lya\ would have been detected in the case that our observations 
would have been deeper, allowing us to
probe greater radii. 
This of course applies to all our galaxies: the regions for which the \lya\ and \ha\ 
fluxes have been integrated is significantly smaller than the total optical extent of the 
galaxies and we may be missing a non-negligible fraction of the diffuse emission.  
The \lya\ escape fractions should in this respect be
seen  as lower limits. On the other hand, some galaxies may
contain star forming regions that are so dusty that not even any \ha\ escapes
in which case the actual \lya\ escape fractions would be lower.

Our modeling software provides us with an estimate of the reddening of the stellar 
continuum below \lya; this is what is required for the subtraction of the continuum from the 
F122M filter. 
However, due to geometrical differences in the configuration of stars and nebular gas, stellar 
and nebular \ebv\ measurements are shown to differ
\citep{Maiz-Apellaniz98,Mas-Hesse_Kunth99,Charlot_Fall00,Calzetti00}. 
Thus we have little information about the reddening in the gas in which \lya\ is formed. 
While differences in the respective reddening measurements can be expected to minimize with 
increasing spatial resolution of the observation, geometrical differences between the stellar 
and nebular components along the line--of--sight will always be present. 
Firstly, we know that \lya\ and Balmer lines must be produced in the same ionized nebulae.
Secondly, we may suspect that due to resonant scattering increasing the traversed path-length, 
\lya\ photons should experience at least the reddening of the Balmer lines, although certain 
geometrical configurations may be envisaged that make this statement uncertain.
Evidently, some circumspection must be exercised when comparing derived reddening with the 
emission (or absorption) of \lya.
With this cautionary remark in mind, we deem it interesting to de-redden the \lya\ fluxes 
(using the reddening found from the SED fit), recompute the \lya\ escape 
fractions, and explore to 
what extent pure dust attenuation can be said to regulate the fraction of escaping
\lya\ photons. In that case we would expect to find  $\mathcal{R}=1$.
After de-reddening we find $\mathcal{R}$ values  between 13\% (\ngc) and 40\% (\eso). 
Hence, even in our dustiest case (NGC\,6090) the observed stellar reddening is 
insufficient in explaining the attenuation of \lya.

\begin{figure*}
\begin{center}
\includegraphics[width=7.5cm]{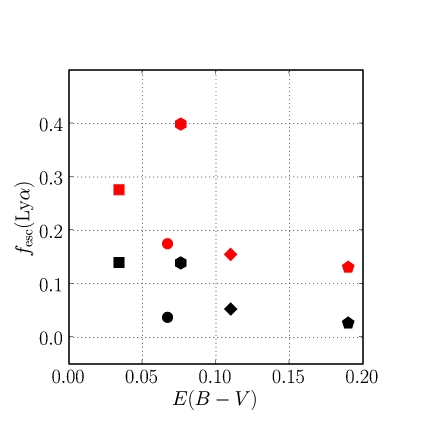}
\includegraphics[width=7.5cm]{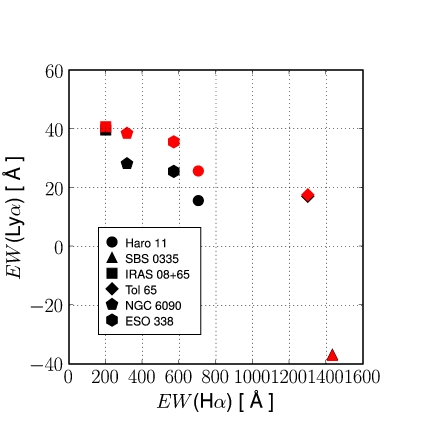}
\end{center}
\caption{Scatter plots for the complete sample: 
\haro\  (circle);
\sbs\   (triangle);
\iras\  (square);
\tol\   (diamond);
\ngc\   (pentagon);
\eso\   (hexagon).
{\em Left}: \lya\ escape fraction reddening {\em vs.} \ebv. 
Black points show the observed escape fractions, and red the escape fractions
corrected for internal extinction. 
Since \sbs\ is a net absorber, it is not shown. 
{\em Right}: $W_{\lyam}$ {\em vs.} $W_{\ham}$. Again, black points show the 
observed values, while for red symbols \lya\ and \ha\ have been corrected for the fitted
\ebv.
\label{fig:globals}
}
\end{figure*}

Figure~\ref{fig:globals} shows where the galaxies fall when $f_{esc,Ly\alpha}$
and $\mathcal{R}$ are compared with \ebv. Any correlation present for 
such a small sample must be viewed upon with skepticism,  but the emitters with 
lowest \ebv\ do have higher  escape fractions.
\eso\ is a dust-poor galaxy with the integrated measurement from this
study finding \ebv$=0.076$. Nevertheless, the applied dust correction  does play a 
significant role, with the same conclusion being reached for \iras, and also perhaps for \haro.
$\mathcal{R}$ in some sense can be thought of as a measure of the
importance of resonant scattering but the effect is so non-linear that quantification is 
not possible.

To strengthen the analysis performed here we would need to independently map the nebular 
extinction, at least mitigating the potential difficulties that arise from blended stars 
and gas.
Such a study is performed at $\sim1$~\arcsec\ resolution in 
\citep{Atek08} 
and we are currently exploring the use of adaptive optics imaging in the 
Br$\gamma$ line for two galaxies.

\subsection{Equivalent widths}

Comparing $W_{\lyam}$ or the \lya\ escape fractions with most of the parameters listed in
Table~\ref{tab:results3} yields almost nothing in the way of trends. 
Indeed this study has never aimed to provide any statistical answers and with
only a sample of six objects could hardly be considered statistically
meaningful. 
Moreover, since the sample was designed to sample a range of starburst
parameters and include emitters and non-emitters of \lya, any trends may purely 
be a feature of the selection. 
That said, the anti-correlation of $W_{\lyam}$ and $W_{\ham}$ mentioned in
Section~\ref{sect:res_globalphot}
is indeed interesting as it seems to be rather tight, although 
this trend is not seen in  the \lya\ escape fraction, with or without
dust correction. 
From the {\em Starburst\,99} models, we estimate that our lowest globally measured
$W_{\ham}$ 
corresponds to an age of around 6~Myr for an instantaneous burst, solar metallicity 
and Salpeter IMF. 
Higher $W_{\ham}$ naturally corresponds to younger ages, with equivalent widths of
several thousand \AA\ expected during the first 3~Myr or so. 
In the studies of 
\cite{Tenorio-Tagle99}
and 
\cite{Mas-Hesse03}
it is concluded that at the youngest times, \lya\ may be expected in absorption,
followed by an emission phase when the mechanical energy return from star
formation has accelerated the ambient medium. 
How the expected equivalent widths of \lya\ and \ha\ co-evolve is
difficult to predict, and  instantaneous burst assumption is certainly an 
oversimplification when dealing with global properties. 
Nevertheless, what we are seeing may be a direct consequence of
evolution.

Related to the \lya\ escape fraction and equivalent width is the escape 
fraction of Lyman
continuum (i.e. ionizing) photons. Some Lyman continuum must escape from 
galaxies at very high-$z$ in order to explain the reionization of the 
universe which apparently was complete by $z\sim 6$ (Fan et al. 2006). 

Escaping Lyman continuum photons
would not contribute to photoionization within galaxies and would thus lower
the luminosities and equivalent widths of \lya\ and other Hydrogen 
recombination lines. However, Lyman continuum leakage would  not necessarily
lower the \lya\ escape fractions or the \lya/\ha\ ratios. On the contrary, 
the existence of \hi\ clear paths through the ISM  would also promote  \lya\ escape.

\cite{Oey07}
studied the ISM in nearby galaxies and their result suggest 
that in starbursts, i.e. galaxies with high \ha\ surface brightness, some of the
Lyman continuum radiation might be leaking out, and there is tentative direct 
detection of Lyman continuum emission in \haro\ (see Sect. \ref{individuals}). 
At high redshift, determining direct escape fractions really means pushing
the observational capabilities, and implied  escape fractions are in general 
low 
\citep{Siana07}, 
although several galaxies have been found with significant 
Lyman continuum leakage 
\citep{Shapley06}.

\section{Perspectives} \label{sect:perspectives}

Not only is \lya\ currently the most favorable window on the galaxy population at
high-$z$ but it is expected to remain so. 
Balmer lines are naturally much easier to interpret than \lya\ but their longer
rest-wavelength makes them less suitable for high-$z$. 
The James Webb Space Telescope promises huge improvements, particularly
regarding \ha\ imaging and spectroscopy at high-$z$ and near- and mid-IR
continuum observations from an environment free from atmospheric background 
emission. 
However, at $z\gtrsim 6$, \ha\ will be redshifted out of the bandpasses of NIRCAM
and NIRSPEC, leaving only MIRI with  limited field--of--view and 
multiplexing capabilities for the detection of \ha. 
Thus especially if \lya\ is, as predicted, visible from beyond the epoch of 
reionization
\citep[e.g.][]{Cen_Haiman00}
then \lya\ is likely to remain one of the primary probes of the high redshift universe 
for the coming decades.

Some rather interesting possible trends have been identified and  
to improve on the study presented here a number of alternatives can be
envisaged. 
Comparison of the co-evolution of $W_{\lyam}$ and $W_{\ham}$ yields results that
are difficult to attach meaning to in a small, hand-picked sample. 
What is required to rectify this is an imaging survey of nearby
star-forming galaxies in \lya, selected in a clean homogeneous manner. 
Some of the targets presented here are of sufficient star-formation rate and FUV
luminosity that they may be directly comparable to some of the objects being
studied by high-$z$ \lya\ observations and such a low-$z$ sample would prove
extremely valuable for co-interpretation of the survey data. 
Indeed, simulations for such a study have already been performed 
(Paper {\sc i}) 
and have demonstrated that results of similar reliability to those presented
here may be obtained with a slightly reduced number of filters.

The use of HST for this project was initially motivated by the need for a FUV-capable
space-based imager to observe \lya\ at $z\sim 0$ and not primarily driven by the
high spatial resolution. 
Given that we see variations in \lya\ between emission and absorption on scales 
down to the resolution limit of the telescope, the need for the high resolving power 
has become essential. 
\hb\ obtained with HST would permit a direct examination of extinction in the
gas phase, an analysis of the role of dust  on a pixel--by--pixel basis, a better 
evaluation of the ionizing photon production, and would also 
enable us to lock the nebular \ebv\ in the continuum fitting software. 
An alternative that we are currently exploring is the use of ground-based
adaptive optics NIR imaging in the Br$\gamma$ line using the NAOS-CONICA
instrument on ESO's VLT.

\section{Conclusions} \label{sect:conclusions}

We have produced continuum subtracted and calibrated \lya\ and \ha\ images for 
a sample of six local star-forming galaxies using the Advanced Camera for 
Surveys onboard Hubble Space Telescope. 
Morphological and photometric analyses have been performed on the \lya, \ha\
and stellar continuum maps. 
In summary:
\begin{itemize}

\item{A variety of morphologies is found, ranging from general absorption to
emission. 
Absorption and emission in individual targets are found to change on scales 
down to the resolution limit of the telescope. 
\lya\ features in emission and absorption show only minor correlation with the 
stellar morphology.  }

\item{Typically \lya\ emission is seen on large scales surrounding the central star
forming regions in the form of low surface brightness halos. 
Moreover, \lya\ is found to be more spatially extended than \ha\ and appears to
emerge almost preferentially at low FUV surface brightnesses. 
At fainter isophotal levels, \lya\ is typically seen in emission, with fluxes
and equivalent widths greater than those that would be predicted from
recombination theory.  }

\item{Observed \lya\ escape fractions range from below 0 (absorber) to about
14\%. Normal corrections for internal dust extinction are insufficient to explain these values.
}

\item{Although any trends found from such a small hand picked sample must be
viewed with caution, the found anti-correlation between the equivalent widths
of \lya\ and \ha\ is quite suggestive. This could be
explained if \lya\ escape is regulated by feedback, i.e. the development of  
wind blown bubbles and ISM flows that allow the photons to avoid the resonant
trapping  by static \hi. }

\item{All the above statements point directly to one conclusion: 
the phenomenon of resonant scattering is observed to be extremely important in 
the regulation of \lya\ emission, and radiative transfer of \lya\ photons in 
star-forming galaxies.  }

\item{A few of our galaxies have far ultraviolet luminosities comparable to some
of the objects being studied through their \lya\ emission at the highest
redshifts. 
Given the low \lya\ escape fractions and complex resonant scattering phenomenon 
under observation here, care should be executed in interpretation of high-$z$ 
survey data.  }

\item{Our sample is fully calibrated and of physical resolution three orders of
magnitude better than the observed candidates at high-$z$. 
It may therefore be valuable in interpretation of high-$z$ datasets and of use to 
compare with the output of radiative transfer codes. 
The images are being released to the community.  }

\end{itemize}

As a service to the community we release calibrated \lya , \ha , and their respective
continuum images to the public domain, for the purpose of modeling high-$z$ galaxies and
\lya\ radiative transfer. The images and documentation are available at 
{\em Centre de Donn{\'e}es astronomiques de Strasbourg (CDS)}
and also
at the following website: 
{\tt http://ttt.astro.su.se/projects/Lyman-alpha/}
Other data products, such as maps of age and reddening and Lyman continuum production 
will be available upon request.

\acknowledgements 

G\"O and MH acknowledge the support from the Swedish National
Space Board (SNSB) and the Research Council (Vetensapsr{\aa}det).
We thank D. Schaerer for useful discussions.

\bibliography{Lyacat_resubm}
\clearpage

\end{document}